\documentclass[apjl]{emulateapj}
\def\ltsima{$\; \buildrel < \over \sim \;$}
\def\simlt{\lower.5ex\hbox{\ltsima}}
\def\gtsima{$\; \buildrel > \over \sim \;$}
\def\simgt{\lower.5ex\hbox{\gtsima}}
\def\kms{km~s$^{-1}$}
\slugcomment{Accepted for publication in ApJ Letters}

\shorttitle{HII regions within a compact high velocity cloud.}
\shortauthors{Bellazzini et al.}

\begin{document}

% 8/12/2014: re-organized including the HVC hypothesis
% 9/12/2014: revised according to Giuseppina comments
% 9/12/2014: revised according to Nicolas comments
% 10/12/2014: adjusted flux units; added flux difference arguments 
%%%%%%%%%%%%%%%%%%%%%%%%%%%%%%%%%%%%%%%%%%%%%%%%%%%%%%%%%%%%%%%%%%%%%
% 15/12/2014: language fixing by Rodrigo; added uncertainties in line ratios 

\title{HII regions within a compact high velocity cloud.\\ 
A nearly star-less dwarf galaxy?$^{\dag}$}

\author{M. Bellazzini\altaffilmark{1}, L. Magrini\altaffilmark{2}, A. Mucciarelli\altaffilmark{3},  G. Beccari\altaffilmark{4}, R. Ibata\altaffilmark{5}, G. Battaglia\altaffilmark{6}, N. Martin\altaffilmark{5,10}, V. Testa\altaffilmark{7}, M. Fumana\altaffilmark{8}, A. Marchetti\altaffilmark{8}, M. Correnti\altaffilmark{9} and F. Fraternali\altaffilmark{3}}
\affil{$^1$ INAF - Osservatorio Astronomico di Bologna, Via Ranzani 1, 40127 Bologna, Italy}
\email{michele.bellazzini@oabo.inaf.it} 
\affil{$^2$INAF - Osservatorio Astrofisico di Arcetri, Largo E. Fermi 5, 50125 Firenze, Italy}
\affil{$^3$Dipartimento di Fisica \& Astronomia, Universit\`a degli Studi di Bologna, 
             Viale Berti Pichat, 6/2, I - 40127 Bologna, Italy}
\affil{$^4$European Southern Observatory, Alonso de Cordova 3107, Vitacura Santiago, Chile}             
\affil{$^5$Obs. astronomique de Strasbourg, Universit\'e de Strasbourg, CNRS, UMR 7550, 11 rue 
             de l'Universit\'e, F-67000 Strasbourg, France }             
\affil{$^6$Instituto de Astrofisica de Canarias, 38205 La Laguna, Tenerife, Spain}             
\affil{$^7$INAF - Osservatorio Astronomico di Roma, via Frascati 33, 00040 Monteporzio, Italy} 
\affil{$^8$INAF - IASF, via E. Bassini 15, 20133, Milano, Italy}            
\affil{$^9$Space Telescope Science Institute, Baltimore, MD 21218}
\affil{$^{10}$Max-Planck-Institut f\"ur Astronomie, K\"onigstuhl 17, D-69117 Heidelberg, Germany}

\altaffiltext{\dag}{Based on data acquired using the Large Binocular Telescope (LBT).
The LBT is an international collaboration among institutions in the United
States, Italy, and Germany. LBT Corporation partners are The University of
Arizona on behalf of the Arizona university system; Istituto Nazionale di
Astrofisica, Italy; LBT Beteiligungsgesellschaft, Germany, representing the
Max-Planck Society, the Astrophysical Institute Potsdam, and Heidelberg
University; The Ohio State University; and The Research Corporation, on behalf
of The University of Notre Dame, University of Minnesota and University of
Virginia.}

\begin{abstract}
Within the SECCO survey we identified a candidate stellar counterpart to the Ultra Compact High Velocity Cloud (UCHVC) HVC274.68+74.70-123, that was suggested by \citet{adams} as a possible mini-halo within the Local Group of galaxies. 
The spectroscopic follow-up of the brightest sources within the candidate reveals the presence of 
two HII regions whose radial velocity is compatible with physical association with the UVHVC.
The available data does not allow us to give a definite answer on the nature of the newly identified system. A few alternative hypotheses are discussed. However, the most likely possibility is that 
we have found a new faint dwarf galaxy residing in the Virgo cluster of galaxies, which we name SECCO~1. Independently of its actual distance, SECCO~1 displays a ratio of neutral hydrogen mass to V luminosity of $M_{HI}/L_V\ga 20$, by far the largest among local dwarfs.
Hence, it appears as a nearly star-less galaxy and it may be an example of the missing links between normal dwarfs and the dark mini halos that are predicted to exist in large numbers according to the currently accepted cosmological model.
\end{abstract}

%% Keywords should appear after the \end{abstract} command. The uncommented
%% example has been keyed in ApJ style. See the instructions to authors
%% for the journal to which you are submitting your paper to determine
%% what keyword punctuation is appropriate.

\keywords{ISM: HII regions --- galaxies: dwarf --- galaxies: star formation}

%% Authors who wish to have the most important objects in their paper
%% linked in the electronic edition to a data center may do so by tagging
%% their objects with \objectname{} or \object{}.  Each macro takes the
%% object name as its required argument. The optional, square-bracket 
%% argument should be used in cases where the data center identification
%% differs from what is to be printed in the paper.  The text appearing 
%% in curly braces is what will appear in print in the published paper. 
%% If the object name is recognized by the data centers, it will be linked
%% in the electronic edition to the object data available at the data centers  
%%
%% Note that for sources with brackets in their names, e.g. [WEG2004] 14h-090,
%% the brackets must be escaped with backslashes when used in the first
%% square-bracket argument, for instance, \object[\[WEG2004\] 14h-090]{90}).
%%  Otherwise, LaTeX will issue an error. 

\section{Introduction}
Modern all-sky surveys have revealed that our census of dwarf galaxies is significantly incomplete even within the  Local Volume \citep[LV, see, e.g.][]{kopo}. The Sloan Digital Sky Survey (SDSS), the PAndAS, and the Pan-STARRS surveys \citep[see, e.g.,][]{belo06,pandas,Pan-STARRS}, among others, have significantly increased the number of known nearby dwarf galaxies, suggesting that many more remain to be discovered \citep{tolle}. Moreover, the newly discovered stellar systems show that actual dwarfs inhabit a much wider range of parameters space than previously believed \citep[see][for recent reviews]{willstra,belo}. For instance, the faintest and most metal-poor star forming galaxies were identified only in recent years, Leo~T \citep{leoT}, lying at $D\simeq 400$~Kpc with $M_V=-8.0$, and Leo~P \citep{leop_1}, having $D\simeq 1.7$~Mpc and $M_V=-9.4$. 
Leo~P, in particular, was discovered with a novel approach, i.e. by identifying a stellar counterpart of an Ultra Compact High Velocity Cloud (UCHVC) identified by the ALFALFA\footnote{\tt http://egg.astro.cornell.edu/alfalfa/} HI survey \citep[see also][for the very recent identification of two dwarfs in the LV as counterparts to compact HI clouds]{tolle2}. Along this line, \citet[][A13 hereafter]{adams} selected from the ALFALFA database a sample of 59 UCHVCs that they proposed as good candidates to be faint (or even star-less) dwarf galaxies associated with low-mass dark matter halos in the distance range 0.25~Mpc$\le D\le$2.0~Mpc. Triggered by the A13 analysis, we started the SECCO survey\footnote{\tt http://www.bo.astro.it/secco/SECCO/} \citep[][B14 hereafter]{secco}, aimed at searching for stellar counterparts in the 25 most promising candidate mini-halos of the A13 sample. 

%------------------------FIG 1-----------------------------------
   \begin{figure*}
   \centering
   \includegraphics[width=17truecm]{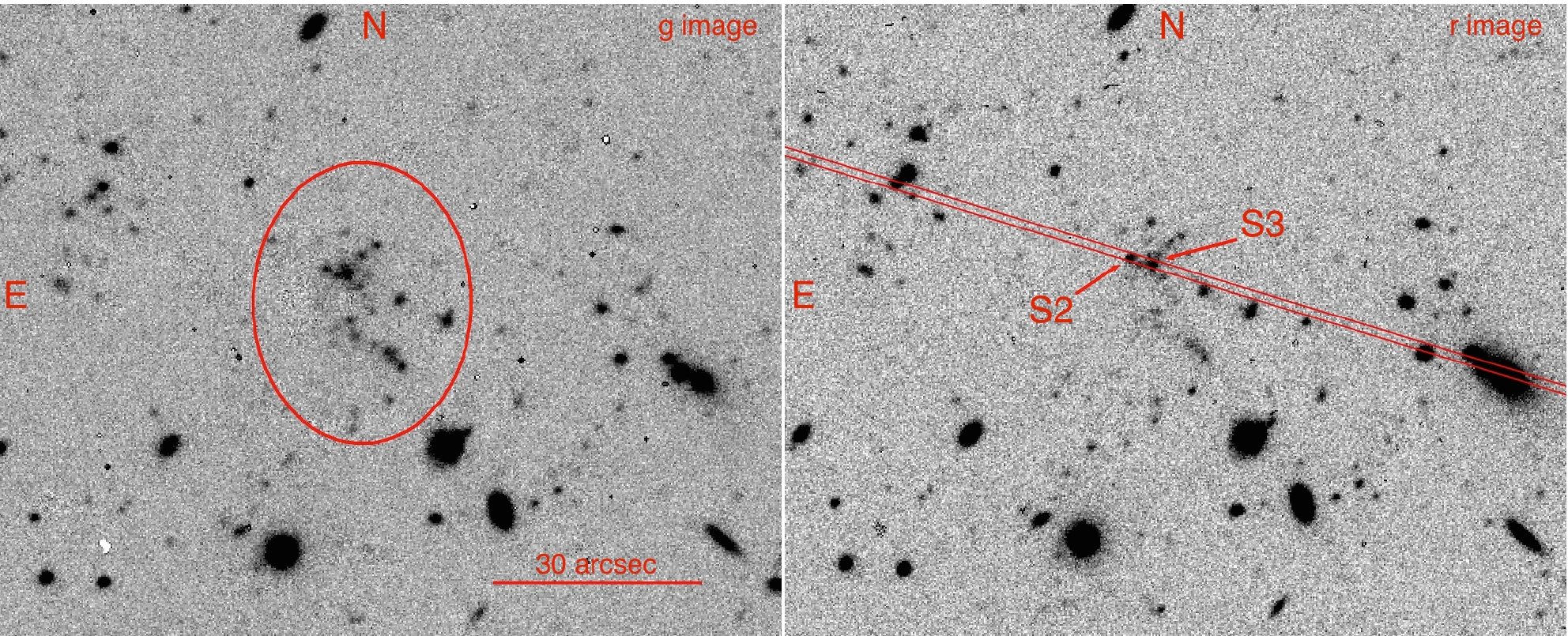}
     \caption{Deep g (left panel) and r (right panel) images zoomed on candidate D1. The ellipse on the left panel has a semi-major axis of 90~px$=20.25\arcsec$ and axis ratio $b/a=0.78$. The half-flux diameters of the associated UCHVC are $4\arcmin\times 5\arcmin$, much larger than the size of this image.
     In the right panel we have highlighted and labelled the two sources for which we obtained MODS1 spectra and the footprint of the adopted slit is also superimposed. 
     %The intensity peaks of S2 and S3 are separated by $3.5\arcsec$.
     }
        \label{targets}
    \end{figure*}
%-------------------------------------------------------------   

Here we present the results of the spectroscopic follow-up of one candidate stellar counterpart, D1, identified in B14 by visual inspection of our deep images. Candidate~D1 appears as an irregular ensemble of partially resolved and relatively bright clumps of compact sources surrounded by some fuzzy light (see Fig.~\ref{targets}) located just $\sim 30\arcsec$ apart from the center of the A13 source HVC274.68+74.70-123 (whose angular size is $5\arcmin \times 4\arcmin$; note that the density profile of the cloud is largely unresolved by the adopted beam). In B14 we concluded that the available data are  consistent with a small dwarf Irregular at $D\ga 3.0$~Mpc
or with a distant group (or cluster) of galaxies (B14). To test these alternative hypotheses we obtained low resolution spectra of the two brightest nearly-point-like sources in the main clump of the candidate, labelled as S2 and S3 in Fig.~\ref{targets}.
We anticipate that the spectrum and the velocity of S3 qualifies it as a stellar counterpart 
of the considered UCHVC, and the most likely hypothesis is that we have found a new very low surface brightness (SB), very star-poor dwarf galaxy in the Virgo cluster of galaxies.

\section{Observations, data reductions and analysis}
\label{obs}

We obtained long-slit spectra with the Multi-Object Double Spectrographs 1 \citep[MODS1;][]{pogge}
mounted on the Large Binocular Telescope (LBT; Mt Graham, AZ) during the nights of June 16, 17, and 19, 2014\footnote{Within an italian Director Discretionary Time programme.}.
A total of seven $t_{exp}=1800$~s exposures were acquired with the MODS1 red-arm equipped with the G670L grism and a slit $1.2\arcsec$ wide. This set-up provides a spectral coverage in the range $\simeq 5000-10000$~\AA ~with a spectral resolution $R=\frac{\lambda}{\Delta\lambda}\sim 1100$.
The spectra were corrected for bias and flat-field, sky-subtracted, wavelength calibrated, then extracted and combined into flux-calibrated summed spectra using the pipeline developed at the {\em Italian LBT Spectroscopic Reduction Center}\footnote{\tt http://lbt-spectro.iasf-milano.inaf.it}.
The targeted sources are very faint ($r\sim 23.0$) and, consequently, the spectra have low signal-to-noise (typically $S/N\sim 5-10$ per pixel, depending on the wavelength), still they are useful for our purposes. 

\begin{figure}
   \centering
   \includegraphics[width=\columnwidth]{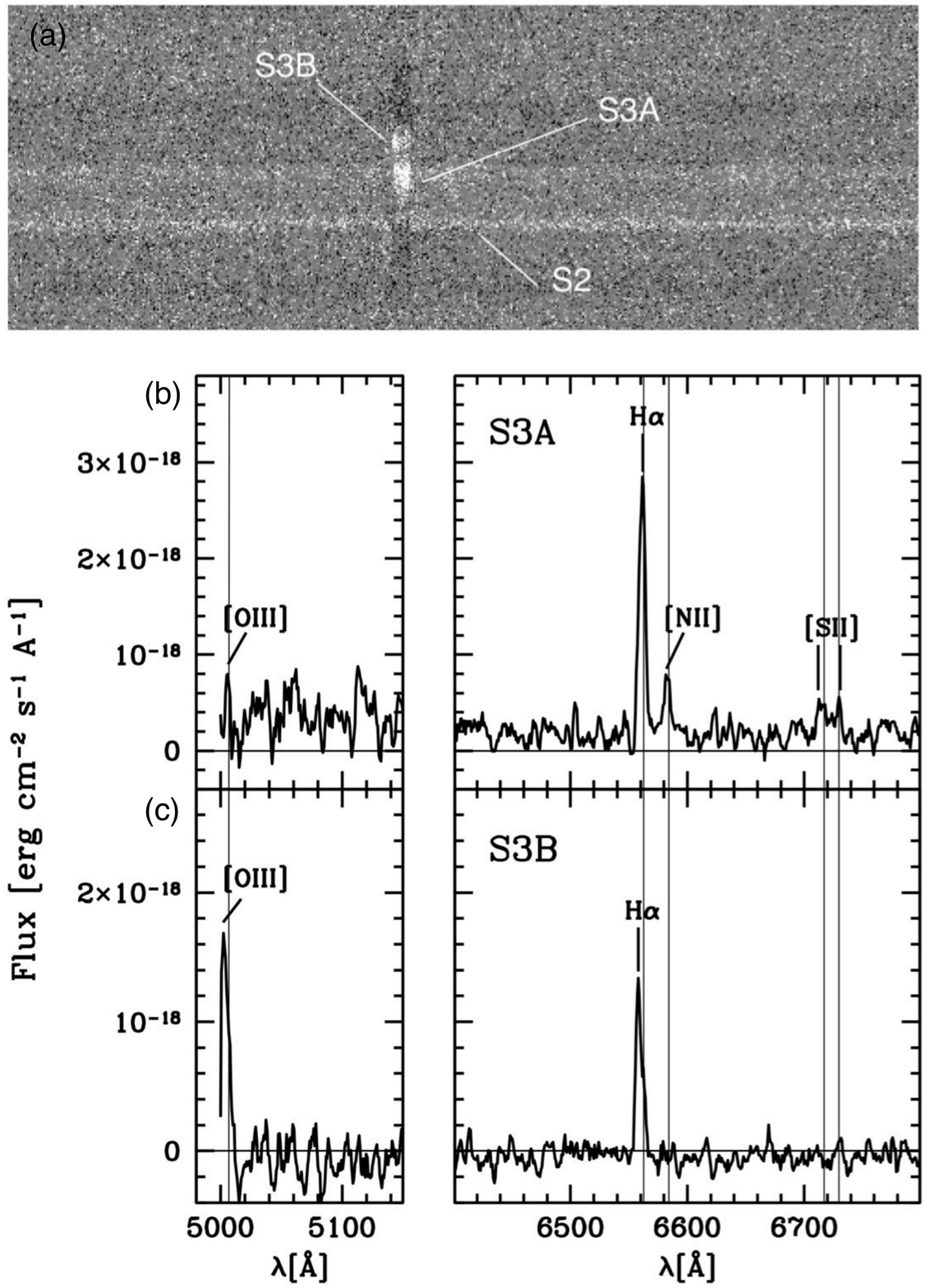}
     \caption{Panel (a): A portion of the combined 2D spectra around the wavelength of rest-frame $H_{\alpha}$. The spectra are labelled after the associated sources. Note the lack of stellar continuum for S3B. Negative spectra footprints (dark features) are artefacts of the process of sky subtraction on stacked spectra.
Panels (b) and (c): Remarkable portions of 1-D spectra of S3A (b) and S3B (c). The spectra have been smoothed with a boxcar filter of size 5~px. The emission lines that have been identified are labelled. Thin vertical lines marks the rest-frame wavelength of the lines, for reference.
          }
        \label{spec}
    \end{figure}

In Fig.~\ref{spec}a we show a portion of the final 2D spectra around the rest-frame H$_{\alpha}$ wavelength.
The stellar continua of S2 and S3 are clearly visible. S3 shows an obvious emission in H$_{\alpha}$ and another
H$_{\alpha}$ emission line, without any detectable stellar continuum, is identified $\sim 2\arcsec$ above (on the opposite side of S2 with respect to S3, along the direction of the slit). We rename the main component of S3 as S3A and the additional continuum-less component S3B (see the labels in Fig.~\ref{spec}a).
The $H_{\alpha}$ emission nearly at rest-frame clearly rules out the hypothesis that candidate~D1 is a distant group/cluster of galaxies. 
The lack of stellar continuum in S3B is likely the main reason why this source cannot be identified in our deep images. However it is clear both from the images and from the spectra that (a) S3A is a complex source, partially resolved and with two additional points sources within $2.2\arcsec$ from its center (B14), and (b) that the spectra of S3A and S3B contaminate each other, i.e. the two sources partially overlap. 
Inspection of the extracted spectra reveals that S2 does not present any spectral feature that can be used to classify or obtain an estimate of its radial velocity. In the following we will discuss in more detail the spectra of the sources S3A and S3B.

%------------------------FIG 3-----------------------------------

%%%%%%%%%%%%%%%%%%%%%%%%%%%%%%%%%%%%%%%%%%%%%%%%%%%%%%%%%%%%%%%%%

In Fig.~\ref{spec}b,c we show a portion of the 1D spectra of S3A and S3B, with the identified spectral lines in emission highlighted and labelled; from left to right [OIII]$_{5007}$, H$_{\alpha}$, [NII]$_{6584}$, [SII]$_{6717}$, and [SII]$_{6730}$. The line ratios of both S3A and S3B (see Table~\ref{targ}) are consistent with those of extragalactic HII regions \citep{kni08}, and not compatible with shocked gas (e.g., supernova remnants). While the [SII/H$_{\alpha}$] and [SII](6717/6731) ratios of S3B lie (formally) in the range typical of Planetary Nebulae (PN), they are still compatible with HII regions if the errors are taken into account. On the other hand S3A falls on the HII regions loci in all the diagnostic diagrams.
The close spatial and kinematic association strongly suggest that S3A and S3B are probably bright knots within an extended HII~ complex with different levels of excitation and displaying variations in the line ratios, like, e.g., the giant HII region NGC~595 in M~33 \citep{relano}. 

Given the low S/N and spectral resolution of our spectra, as well as the paucity of spectral features, we estimate the radial velocity in different ways, using original or smoothed (with a 5~px wide boxcar filter)  spectra, including and excluding one or more lines from the analysis, and in all cases we obtain consistent results. Our final best-estimate of the {\rm heliocentric radial velocity} ($V_h$ in the following) of S3A is obtained by cross-correlation of the smoothed spectrum with a synthetic flat spectrum with four gaussians of the proper width added at the rest-frame wavelength of the four lines identified in the right panel of Fig.~\ref{spec}b. In this way we get $V_h=-77 \pm 42$~\kms. The attached error bar takes into account both statistic and systematic uncertainties affecting our estimate, including significant uncertainties in the velocity zero point (VZP).  Unfortunately, the low S/N of our spectra prevented the possibility to correct the VZP using telluric features. Since we placed the intensity peak of S3A (as identified in B14) at the center of the slit, the error on the VZP due to imperfect alignment should be limited to $\pm 0.5$~px, corresponding to $\pm 40$~\kms, by far the dominant source of uncertainty.
On the other hand we had no previous knowledge of the position of S3B and its luminosity peak can be misaligned with respect 
to the slit center by as much as $\simeq 4.9$~px, corresponding to systematics in $V_h$ up to $\sim 200$~\kms; we adopt half of this value as a rough estimate of the $1\sigma$ error on the velocity of S3B.
Using only $H_{\alpha}$, de-blended from the contamination by S3A, we find $V_h=-260 \pm 100$~\kms~ from the smoothed spectra of S3B.

Given their proximity in the sky ($\sim 2\arcsec$) and in velocity, it seems exceedingly unlikely that S3A and S3B are not physically associated. Hence 
the velocity of S3B is probably affected by a significant mis-centering bias and, consequently, we adopt the velocity of S3A as the systemic velocity of the two sources. This velocity is compatible with the velocity of the associated UCHVC ($V_h=-128 \pm 6$ \kms, from A13), {\em implying that at the center of HVC274.68+74.70-123 there is indeed a stellar counterpart, with ongoing star formation}. This conclusion is also supported by the fact that the stellar clump including S2 and S3, as well as another stellar blob associated with candidate D1 a few arcsec to the South of S3, are also detected in NUV images from the Galaxy Evolution Explorer (GALEX) mission, and measured as a single source with $NUV_{mag}=20.84\pm 0.10$. Unfortunately, no FUV images of D1 are available. 

With the information available in the spectra, we can attempt to
estimate some physical and chemical properties of the two regions. 
From the ratio  [SII]6717/[SII]6730 we obtain an electron density 
$\sim 1100$~cm$^{-3}$ for S3A (assuming an electron temperature 10000~K).
From the line ratio N2=log([NII]$_{6584}$/H$_{\alpha}$)=$-0.52\pm 0.06$, computed
from the cumulative spectrum (S3A+S3B), we obtain a rough estimate of the average
metallicity of the whole HII~ complex. Using the calibration for HII~ regions by \citet{petti} we find an oxygen abundance 12+log(O/H)=$8.6\pm 0.5$, larger than the typical values found in HII~ regions of local dwarfs \citep{lee}.

\section{A cloud in the Galactic Halo?}
\label{hvc}

If S3A is a genuine HII region, the possibility that it belongs to our own Galaxy seems highly unlikely. Assuming that the most compact HII~regions with the
electron density of S3A (see below) have diameters $\ge 1$~pc \citep{hunt}, and adopting, very conservatively, the observed FWHM  ($\simeq 1.1\arcsec$ on the r images, where the stellar Point Spread Function  has FWHM$\simeq 0.95\arcsec$) as the angular size of the region, it turns out that S3A must lie at least at $D\ga 200$~kpc from us ($D\ga 100$~kpc if we consider the separation between S3A and S3B as the typical size of the complex). Moreover, Galactic star forming regions should lie in the Galactic disc, while the direction to HVC274.68+74.70-123 is nearly perpendicular to it and its velocity is completely incompatible with the disc. 
The hypothesis of a HVC located in the outer Galactic halo ($D\ga 100$~kpc) and including a star forming region appears similarly unlikely, since the cases of Galactic HVCs with an associated stellar component are exceedingly rare, if any \citep[][]{ivechrist,stark}.

There is another possibility that may be considered. HVCs are known to host ionized gas, in addition to HI \citep[see, e.g.,][and references therein]{put}. It can be conceived that a relatively nearby HVC, partially ionised by radiation from massive stars in the Galactic disc, can be closely aligned along the line of sight, by chance, with background unrelated sources identified as candidate D1.
In this case the emission lines we see in the spectra of S3A and B would not be associated with these sources but with the cloud,
and would be superimposed on the featureless stellar continuum of the background sources. The main argument against this scenario is that emission lines are not seen superimposed on the continuum of S2, lying just $\simeq 3.5\arcsec$ apart from S3A, while H$_{\alpha}$ emission in HVCs is observed over scales comparable with the dimension of the clouds \citep[see, e.g.][]{haf}. We note that no other rest-frame H$_{\alpha}$ emission is detected along the $5.0\arcmin$ long MODS slit, in our spectra, suggesting that the observed emission lines are indeed associated with the S3 stellar source. Moreover, HVCs display typical H$_{\alpha}$ fluxes of $\simeq 0.1$~Rayleigh and generally lower than $\simeq 0.4$~Rayleigh \citep{put}, while, assuming an area of 1~arcsec$^2$ for S3A and B, their H$_{\alpha}$ fluxes are  $\simeq 3.3$~Rayleigh and $\simeq 1.5$~Rayleigh, i.e. a factor from $\sim 3$ to $\ga 10$ larger than this.

The case for a Galactic HVC (associated to or simply aligned with S3A and B), as unlikely as it may appear, would provide a natural explanation (a) for the radial velocity of the system, that is clearly incompatible with the Hubble flow, and (b) for the derived N2 value, that is fully compatible with those observed in HVCs \citep{put}, while, in the hypothesis of a dwarf galaxy (explored in the next section) it would imply an exceedingly high metallicity for the estimated optical luminosity.

\section{A nearly star-less dwarf galaxy?}
\label{dwarf}

Let us now consider the hypothesis that we have discovered a new faint dwarf galaxy,
where a few clumps of star forming regions are enclosed within a (comparatively) huge HI~ cloud. We preliminarily name it SECCO~1, after the name of the survey.
Since there is no detectable smooth distribution of light around the clumps, to get an estimate of the integrated magnitude
and of the half-light radius we recur to simple aperture photometry, taking the ellipse plotted in the left panel of Fig.~\ref{targets}
as the tentative outer limit of the stellar system. The results are reported in Tab.~\ref{secco1}. 
The integrated color, $(g-r)_0=0.27$, is typical of a star-forming galaxy (see Table ~\ref{targ}). 
Unfortunately we have only broad constraints on the distance. The fact that we do not resolve RGB stars implies that $D\ga 3$~Mpc (see B14), if an old population is there. 
In any case, the system is not participating into the Hubble Flow, since for $D\ga 3$~Mpc this implies $V_h\ga +150$~\kms \citep[][MC12 hereafter]{mcc}. 

\begin{figure}
   \centering
   \includegraphics[width=\columnwidth]{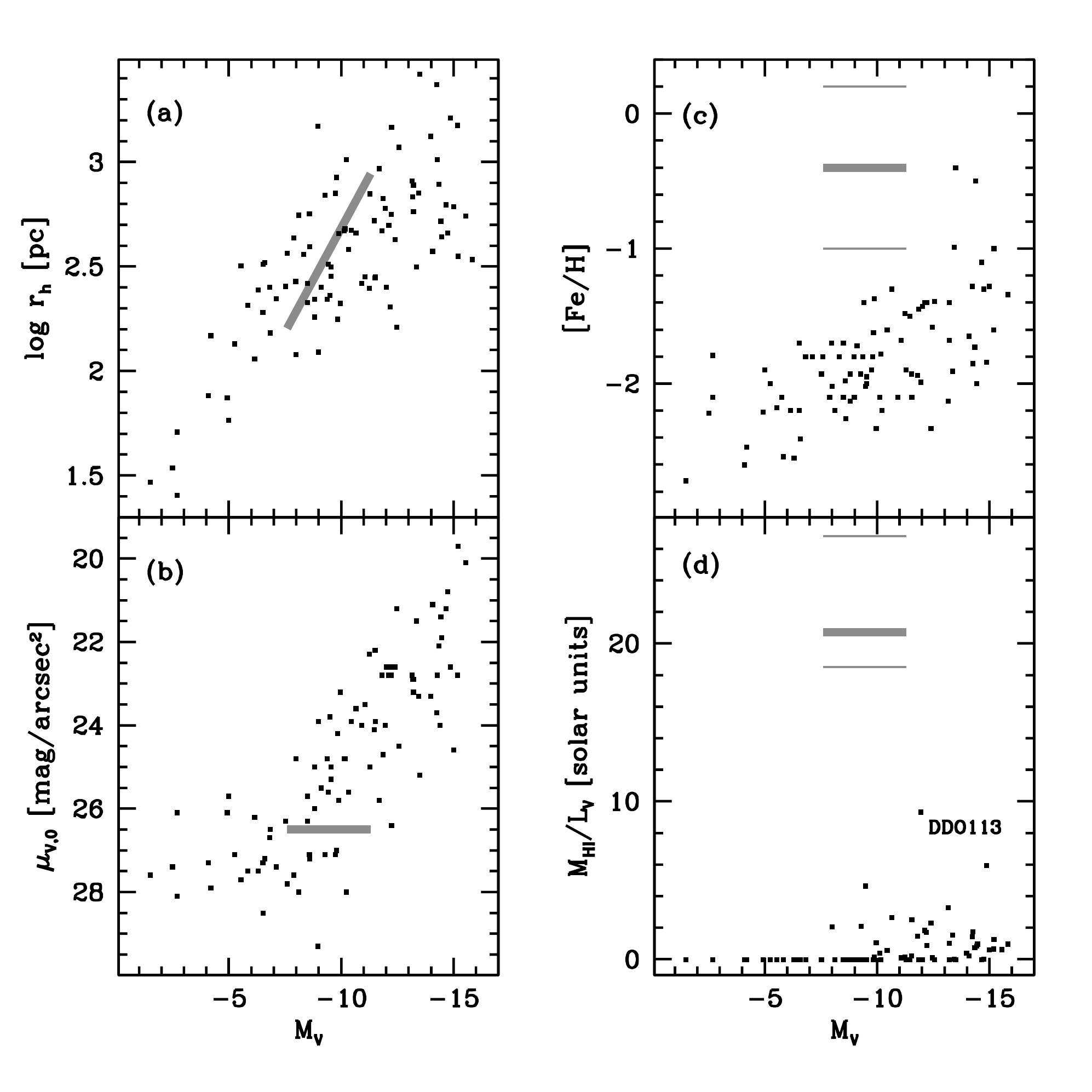}
     \caption{The range in various physical parameters spanned by SECCO~1, depending on the assumed distance (from D=3.0~Mpc to D=16.5~Mpc, grey bars), is compared with the distribution of dwarf galaxies in the LV from the MC12 catalog (black dots). The grey lines in panels c and d enclose the $\pm 1\sigma$ error interval in [Fe/H] and $M_{HI}/L_V$, respectively. 
     }
        \label{compa}
    \end{figure}

Adopting distances in the range 3.0~Mpc$\le D\le$16.5~Mpc \citep[the latter value corresponding to the distance of the Virgo cluster,][]{mei} we can derive several physical characteristics of the stellar and gaseous body of SECCO~1\footnote{HI properties and dynamical mass are derived from Eqs. 7, 8, and 9 of A13; note that the dynamical mass is estimated from the HI velocity width, hence it is very uncertain}. 
In this distance interval, the system nicely fits into the observed $M_V-r_h$ and $M_V-\mu_V$ relations for local dwarf galaxies shown by MC12 (see Fig.~\ref{compa}a,b). The relevant quantities span the ranges $-7.6\la M_V\la -11.3$, 160~pc$\la r_h\la$880~pc, $10^{6.3}~M_{\sun}\la M_{HI}\la 10^{7.8}~M_{\sun}$, and $10^{8.4}~M_{\sun}\la M_{dyn}\la 10^{9.1}~M_{\sun}$. For $D\le 16.5$~Mpc the Star Formation Rate, estimated from the H$_{\alpha}$ luminosity according to \citet{kenni}, is $\le 6.1\times 10^{-6}$~M$_{\sun}$~yr$^{-1}$, among the lowest rates recorded in local star-forming dwarfs \citep[see, e.g.,][]{J14}. 

%------------------------FIG 4-----------------------------------

%%%%%%%%%%%%%%%%%%%%%%%%%%%%%%%%%%%%%%%%%%%%%%%%%%%%%%%%%%%%%%%%%

Within this framework we see only two reasonable hypotheses for the location of SECCO~1.

\noindent
1. SECCO~1 is a very low luminosity, low SB analog of the low-mass isolated star forming galaxies studied by \citet{cannon}. These lie in the distance range 3.2~Mpc$\le D\le$18.7~Mpc and are associated with ALFALFA small clouds listed by \citet{AGC} which have stellar counterparts that are visible in SDSS images. HVC274.68+74.70-123 corresponds to AGC226067 in the \citet{AGC} catalog, where it is commented with the note ``No identifiable optical counterpart''. Indeed SECCO~1 is barely visible even in our images, more than 4~mag deeper than SDSS. Hence it would represent an extremely dark member of this population.
Still, all of the \citet{cannon} galaxies are more or less compatible with being participants in the local Hubble Flow, while SECCO~1 is not. 

\noindent
2. SECCO~1 is a very low luminosity, low SB member of the Virgo cluster. Indeed, it lies at just $2.44\degr$ from the center of the cluster and has a velocity well within the range spanned by cluster members \citep[that reaches $V_h<-600$~\kms,][]{gira}. This hypothesis also implies the lowest (and most reasonable) total mass to light ratio for the system, $M_{dyn}/L_V\simeq 470$, and the highest baryon fraction, $f_b=0.046$. 

Hypothesis 2 appears as the more natural explanation of the available observational material. Still, according to the stellar mass - metallicity relation for galaxies by \citet{ellis}, the metallicity we obtain from N2 implies a stellar mass of about $10^8-10^9~M_{\sun}$ for SECCO~1, which appears incompatible with the estimates from the integrated magnitude, independently of the assumed distance ($M_{*}\la 10^{6.5}~M_{\sun}$, adopting $M/L_V=1.0$). However the $1\sigma$ lower limit of our [Fe/H] estimate is marginally compatible with the $M_V$~-~[Fe/H] relation for LV dwarfs by MC12, if it is assumed that SECCO~1 lies in Virgo (see Fig.~\ref{compa}c).\footnote{The size, the integrated magnitude, the line ratios and the metallicity of SECCO~1, if located in Virgo, are also compatible with those of ``fireballs'', star-forming knots
of gas ram-pressure-stripped from dwarf galaxies within galaxy clusters \citep[see][]{yoshi,fuma}. However, 
the lack of any obvious parent galaxy, the much lower star formation rate, and the large size of the associated UCHVC
militates against the possibility of SECCO~1 being a left-over fireball. }

The final word on the nature of SECCO~1 can be obtained only with Hubble Space Telescope observations, that would allow one to resolve RGB stars (if any) even at the distance of Virgo in such a low SB system \citep{jang}. High spatial resolution HI observations and H$_{\alpha}$ imaging may also provide very useful insight.
 
If SECCO~1 is indeed a dwarf galaxy, it would be an extremely gas-dominated and star-poor one, akin to the isolated 
``almost dark'' dwarf galaxies discussed by \citet{candark}. Independent of the assumed distance, its $M_{HI}/L_V$ (and $M_{HI}/L_r$) ratio, in solar units, is $\ga 20$, much larger than in the dwarfs listed by MC12 (Fig.~\ref{compa}d) and by \citet{cannon}, 
that have $M_{HI}/L_V\la 10$. Systems with $M_{HI}/L_V\ga 20$  are quite rare 
\citep{maddox,candark} 
and may constitute the tip of the iceberg of a population of completely dark mini-haloes that is predicted to exist by  the $\Lambda$-CDM cosmological model \citep{ricotti,sawa}.

\acknowledgments
We acknowledge the support from the LBT-Italian Coordination Facility and the Italian LBT Spectroscopic Reduction Center for the execution of observations, data distribution and reduction. We are grateful to S. Bardelli, A. Bragaglia, A. Buzzoni, A. McConnachie, M. Mignoli, and P. Montegriffo for useful discussions and/or assistance. We are grateful to an anonymous referee
for her/his insightful suggestions.
M.B and F.F. acknowledge the support from PRIN MIUR 2010-2011 project ``The
Chemical and Dynamical Evolution of the Milky Way and Local Group Galaxies'',
prot. 2010LY5N2T. 
G. Battaglia gratefully acknowledges support through a Marie-Curie action Intra European Fellowship, funded from the European Union Seventh Framework Program (FP7/2007-2013) under Grant agreement number PIEF-GA-2010-274151, as well as the financial support by the Spanish Ministry of Economy 
and Competitiveness (MINECO) under the Ram\'on y Cajal Program 
(RYC-2012-11537).
This research has made use of the SIMBAD database, operated at CDS, Strasbourg, France.
This research has made use of the NASA/IPAC Extragalactic Database (NED) which is operated by the Jet Propulsion Laboratory, California Institute of Technology, under contract with the National Aeronautics and Space Administration.

{\it Facilities:} \facility{LBT}, \facility{GALEX}.

\clearpage

%%%%%%%%%%%%%%%%%%%%%%%%%%%%%%%%%%%%%%%%%%%%%%%%%%%%%%%%%%%%%%%%%%%%%%%%%%%%%%%%%%%%%%%%%%%%%%%%%%%%%%%%%%%%
\begin{table}
  %\begin{center}
  \caption{Spectroscopic targets observed with MODS1}
  \label{targ}
  \begin{tabular}{lcccc}
\tableline	      
           &  S2           &     S3A                 &  S3B    & units\\
\tableline	      
RA$_{J2000}$ &  12:21:54.4   &  12:21:54.2              & \nodata                  & hh:mm:ss    \\   
Dec$_{J2000}$& +13:27:42.6   & +13:27:41.6              & \nodata                  & dd:mm:ss   \\   
$g_0$        &$24.12\pm 0.07$& $23.63\pm 0.06^a$        & \nodata                  & mag \\
$r_0$        &$23.07\pm 0.04$& $23.07\pm 0.05^a$        & \nodata                  & mag \\
$V_h$        & \nodata       & $-77\pm 42$              & $-260 \pm100^b$                 & km~s$^{-1}$ \\
$[OIII]$ flux    & \nodata & $2.0\pm 1.0\times 10^{-18}$ & $11.0\pm 1.0\times 10^{-18}$& erg cm$^{-2}$ s$^{-1}$\\
H$_{\alpha}$ flux  & \nodata& $19.0\pm 1.0\times 10^{-18}$& $8.5\pm1\times 10^{-18}$    & erg cm$^{-2}$ s$^{-1}$\\
$[NII]$ flux	      & \nodata& $6.0\pm 1\times 10^{-18}$   & $\le 2.3\pm1.4\times 10^{-18}$$^c$& erg cm$^{-2}$ s$^{-1}$\\
$[SII]_{6717}$ flux& \nodata& $2.7\pm 1\times 10^{-18}$& $\le 0.3\pm1.4\times 10^{-18}$$^c$& erg cm$^{-2}$ s$^{-1}$\\
$[SII]_{6730}$ flux& \nodata  & $2.9\pm 1\times 10^{-18}$& $\le 0.4\pm1.4\times 10^{-18}$$^c$& erg cm$^{-2}$ s$^{-1}$\\
\tableline	      
\end{tabular} 
\tablecomments{All data from B14 except for the heliocentric radial velocity and the line fluxes. 
$^a$ To account for the complex and extended nature of S3A we have also performed aperture photometry with aperture radius 
of 10~px, finding $g=22.57 \pm 0.06$ and $r=22.37 \pm 0.05$, and, consequently, $(g-r)_0=0.20\pm 0.08$. 
$^b$ See text for discussion on velocity uncertainties.} $^c$ Upper limits derived by subtracting S3A fluxes from those obtained from the combined S3A+S3B spectrum.
%\end{center}
\end{table}
%%%%%%%%%%%%%%%%%%%%%%%%%%%%%%%%%%%%%%%%%%%%%%%%%%%%%%%%%%%%%%%%%%%%%%%%%%%%%%%%%%%%%%%%%%%%%%%%%%%%%%%%%%%%

%\clearpage

%%%%%%%%%%%%%%%%%%%%%%%%%%%%%%%%%%%%%%%%%%%%%%%%%%%%%%%%%%%%%%%%%%%%%%%%%%%%%%%%%%%%%%%%%%%%%%%%%%%%%%%%%%%%
\begin{table}
  %\begin{center}
  \caption{Properties of SECCO~1.}
  \label{secco1}
  \begin{tabular}{lc}
\tableline	      
RA$_{J2000}$  &  12:21:54 \\
Dec$_{J2000}$ & +13:27:36 \\
l             &  274.680$\degr$ \\
b             &   74.688$\degr$ \\
$g_{int}$     & $20.1\pm 0.2$ \\
$r_{int}$     & $19.8\pm 0.2$ \\
 E(B-V)       & 0.048 \\
$V_{int}$     & $19.9\pm 0.2$ \\
$r_h$         &$\sim 11\arcsec$ \\
$\mu_{e,V}$   &$\simeq 27.0$~mag~arcsec$^{-2}$ \\
%$\mu_{0,V}$   &$\sim 26.5$~mag~arcsec$^{-2}$ \\
$V_h$         &$-128\pm 6$ ~km~s$^{-1}$\\
\tableline	      
\end{tabular} 
\tablecomments{Coordinates and E(B-V) from B14. 
$V_h$ is from A13. $\mu_{e,V}$ is the mean V-band SB within the half-light radius.}
%\end{center}
\end{table}
%The integrated V magnitude has been obtained from the g and r magnitudes using the transformations by Lupton (2005; see {\tt http://classic.sdss.org/dr7/algorithms/sdssUBVRITransform.html}). 

%%%%%%%%%%%%%%%%%%%%%%%%%%%%%%%%%%%%%%%%%%%%%%%%%%%%%%%%%%%%%%%%%%%%%%%%%%%%%%%%%%%%%%%%%%%%%%%%%%%%%%%%%%%%

\end{document}